\documentclass[lettersize,journal]{IEEEtran}
\usepackage[nolist]{acronym}
\usepackage{siunitx}

\usepackage{bm}

\usepackage{url}
\usepackage{color}
\usepackage{amsmath,amsfonts}
\usepackage{algorithmic}
\usepackage{algorithm}
\usepackage{array}
\usepackage[caption=false,font=normalsize,labelfont=sf,textfont=sf]{subfig}
\usepackage{textcomp}
\usepackage{stfloats}
\usepackage{url}
\usepackage{verbatim}
\usepackage{graphicx}
\usepackage{cite}
\PassOptionsToPackage{hyphens}{url}\usepackage{hyperref}
\usepackage[sort,compress,capitalise]{cleveref}	

\newcommand{\cs}[3]{#1_{\text{#2}}^{\text{#3}}}
\newcommand{\sref}[1]{(\ref{#1})}
\newcommand{\COtwoeq}{CO\textsubscript{2}\,eq}
\DeclareSIUnit{\univcurr}{\text{\textcurrency}}

\begin{acronym}
    \acro{DCO}{Data Center Operator}
    \acro{ADN}{Active Distribution Network}
    \acro{CVaR}{Conditional Value-at-Risk}
    \acro{DER}{Distributed Energy Resource}
    \acro{ORC}{Organic Rankine Cycle}
    \acro{TSO}{Transmission System Operator}
    \acro{DSO}{Distribution System Operator}
    \acro{BESS}{Battery Energy Storage System}
    \acro{PV}{PhotoVoltaic}
    \acro{CPU}{Central Processing Unit}
    \acro{GPU}{Graphics Processing Unit}
    \acro{QoS}{Quality of Service}
    \acro{SLA}{Service Level Agreement}
    \acro{GCP}{Grid Connection Point}
    \acro{SOS2}{Special Ordered Set of type 2}
    \acro{GBR}{Gradient Boost Regression}
    \acro{PUE}{Power Usage Effectiveness}
    \acro{VCC}{Virtual Capacity Curve}
    \acro{MILP}{Mixed Integer Linear Program}
    \acro{PDF}{Probability Distribution Function}
    \acro{DCO}{Data Center Operator}
    \acro{ETS}{Emissions Trading System}
    \acro{ToU}{Time-of-Use}
    \acro{AI}{Artificial Intelligence}
    \acro{IEA}{International Energy Agency}
    \acro{DLC}{Direct Liquid Cooling}
    \acro{CC}{Chance Constraint}
    \acro{DH}{District Heating}
    \acro{WL}{workload}
    \acro{GHI}{Global Horizontal Irradiance}
    \acro{PPA}{Power Purchase Agreement}
    \acro{KDE}{Kernel Density Estimation}
\end{acronym}

\begin{document}

\title{Should Small-Scale Data Centers Participate in the Day-Ahead Electricity Market?}

\author{
\IEEEauthorblockN{Enea Figini,~\IEEEmembership{Student Member,~IEEE} and Mario Paolone,~\IEEEmembership{Fellow,~IEEE}}
\thanks{The authors are with the Swiss Federal Institute of Technology (EPFL) in 1015 Lausanne, Switzerland (e-mail: \{enea.figini,~mario.paolone\}@epfl.ch).}
\thanks{Preprint: 4th of May 2026.}}

\markboth{PREPRINT}%
{Shell \MakeLowercase{\textit{et al.}}: A Preprint Using IEEEtran.cls}


\maketitle

\begin{abstract}
The global race to artificial intelligence competitive advantage is challenging electricity grids by demanding growing data center capacity. Addressing this challenge requires synergistic operational strategies that integrate data centers into electricity markets while supporting grid operation. This work proposes a bilateral power purchase agreement between small-scale data centers and distribution system operators, enabling data center participation in the day-ahead electricity market. To facilitate market participation, we develop a scenario-based, risk-averse bidding strategy that leverages flexibility from local energy resources, waste heat recovery, and data center workload. The strategy jointly minimizes operational costs and carbon emissions, creating a carbon-aware cost-effective framework for data center integration in the electricity day-ahead market. The method is evaluated on a study case comparing a conventional time-of-use supply scheme with the proposed custom power purchase agreement, showing a potential 22\% cost reduction, thus highlighting financial opportunities for small-scale data centers day-ahead electricity market participation. Two additional case studies illustrate the marginal effects of: (i) data center flexible workload on energy costs and (ii) virtual de-rating of grid transfer capacity.
\end{abstract}


\section{Nomenclature}
\begin{IEEEdescription}[\IEEEusemathlabelsep\IEEEsetlabelwidth{$\cs{E}{bess}{min}$, $\cs{E}{bess}{max}$}]
\setlength{\itemsep}{2pt}
\item[\textbf{Sets}]
\item[$\mathcal{D}$] Set of timesteps in a day.
\item[$\mathcal{W}$] Set of timesteps in a week.
\item[$\mathcal{T}$] Set of timesteps in the bidding problem.
\item[$\Omega$] Set of scenarios in the bidding problem.
\item[$\mathcal{C}$] Set of clusters in the data center.
\item[$\mathcal{R}$] Set of computational resources in the data center.
\item[$\mathcal{R}_{\text{mem}}$] Set of memory resources.
\item[$\mathcal{R}_\text{cooled}$] Set of resources with Direct Liquid Cooling (DLC).
\item[\textbf{Indices}]
\item[$t$] Timestep index.
\item[$\omega$] Scenario index.
\item[c] Data center cluster identifier.
\item[res] Resource identifier.
\item[$i$] Sampling index for the Organic Rankine Cycle (ORC) heat to power piecewise approximation.
\item[\textbf{Parameters}]
\item[$N_s$] Number of samples for the linear interpolation of the input heat-to-output ORC power curve.
\item[$\cs{P}{gcp}{rated}$] Rated power of the Grid Connection Point (GCP) connection.
\item[$\cs{P}{gcp,cap}{$t$}$] Virtual transfer capacity at the GCP.
\item[$\Delta \cs{P}{gcp}{$t$}$] GCP de-rating amount.
\item[$\cs{P}{gcp,min}{}$] Guaranteed transfer capacity at the GCP.
\item[$\cs{t}{daily,lim}{}$] Daily maximum duration of GCP de-rating.
\item[$\cs{t}{weekly,lim}{}$] Weekly maximum duration of GCP de-rating.
\item[$\Delta T$] Time interval between timesteps in the bidding problem.
\item[$\alpha$] Conditional Value at Risk (CVaR) confidence level in the bidding problem.
\item[$\beta$] CVaR weight in the bidding problem objective.
\item[$\pi_\text{carbon}$] Carbon price.
\item[$\cs{\pi}{spot}{$\omega,t$}$] Electricity spot price.
\item[$\cs{\pi}{short}{$\omega,t$}$] Short imbalance price.
\item[$\cs{\pi}{long}{$\omega,t$}$] Long imbalance price.
\item[$\cs{\pi}{heat}{}$] Heat selling price.
\item[$S_\text{cap}$] Renewable share target.
\item[$\alpha_r$] Allowed fraction of violations of $S_\text{cap}$.
\item[$\kappa^{c,\text{res}}$] Cluster capacity of given resource.
\item[$\Gamma_\text{inelastic}$] Guaranteed fraction of  served inelastic demand.
\item[$\Gamma_\text{flexible}$] Guaranteed fraction of served flexible demand.
\item[$\gamma_\text{mem}^{c,\text{res}}$] Historical average of the memory-to-compute ratio.
\item[$\eta_\text{PUE}$] Data center power usage effectiveness.
\item[$\cs{\rho}{inter}{c}$] Intercept of the linear regression between resource usage and IT power consumption.
\item[$\cs{\rho}{coeff}{c,\text{res}}$] Affine coefficient of the linear regression between resource usage and IT power consumption.
\item[$\rho_\text{cooled,idle}^\text{c}$] Cluster idle power consumption of resources with on-chip micro-fluidic cooling. 
\item[$\eta_\text{rec}^{c,\text{res}}$] Heat recovery efficiency.
\item[$(\dot{q}_i,p_i)$] ORC heat-to-power samples.
\item[$\eta_\text{bess}$] Battery Energy Storage System (BESS) one-way efficiency.
\item[$\cs{E}{bess}{min}$, $\cs{E}{bess}{max}$] Energy bounds for BESS operation.
\item[$\cs{P}{bess}{rated}$] Rated BESS power.
\item[$\cs{L}{bess}{cycles}$] Rated cycles of BESS.
\item[$\cs{E}{bess}{rated}$] Rated capacity of BESS.
\item[$\cs{\Pi}{bess}{investment}$] Investment cost of BESS.
\item[$\cs{C}{bess}{LCA}$] Life-cycle emissions of BESS.
\item[$\cs{P}{pv}{rated}$] Rated PV power.
\item[$\cs{i}{ghi}{$\omega,t$}$] Forecasted Global Horizontal Irradiance (GHI).
\item[$\cs{i}{ghi,ref}{}$] Reference GHI.
\item[$\cs{c}{gcp}{$\omega,t$}$] Grid carbon intensity of electricity.
\item[$\cs{s}{gcp}{$\omega,t$}$] Grid renewable share of electricity.
\item[$\cs{\dot{Q}}{demand}{$\omega,t$}$] District heating demand forecast.
\item[$\cs{I}{}{$\omega,t,c,\text{res}$}$] Inelastic workload (WL) demand forecast.
\item[$\cs{F}{}{$\omega,c,\text{res}$}$] Flexible WL demand forecast.
\item[\textbf{Variables}]
\item[$\bm{\Pi}_\text{op}^{\omega}$] Total daily operational cost.
\item[$\cs{\bm{\Pi}}{d-a}{$\omega,t$}$] Day-ahead electricity supply cost.
\item[$\cs{\bm{\Pi}}{imb}{$\omega,t$}$] Imbalance settlement cost.
\item[$\cs{\bm{\Pi}}{op,DERs}{$\omega,t$}$] Distributed Energy Resources (DERs) operational cost.
\item[$\cs{\bm{\Pi}}{heat}{$\omega,t$}$] Heat revenue.
\item[$\cs{\bm{C}}{gcp}{$\omega,t$}$] Electricity imports equivalent carbon emissions.
\item[$\cs{\bm{C}}{op,DERs}{$\omega,t$}$] DER operational equivalent carbon emissions.
\item[$\cs{\bm{u}}{}{$\omega,t,c,\text{res}$}$] Data center resource usage.
\item[$\cs{\kappa}{v}{$t,c,\text{res}$}$] Virtual capacity of resource.
\item[$\cs{\bm{P}}{dc}{$\omega,t$}$] Data center power consumption.
\item[$\cs{\bm{\dot{Q}}}{rec}{$\omega,t$}$] Total recovered heat.
\item[$\cs{\bm{\dot{Q}}}{orc,in}{$\omega,t$}$] Recovered heat sent to ORC.
\item[$\cs{\bm{\dot{Q}}}{sold}{$\omega,t$}$] Recovered heat sold to district heating.
\item[$\cs{\bm{\dot{Q}}}{lost}{$\omega,t$}$] Recovered heat lost.
\item[$\cs{\bm{P}}{orc}{$\omega,t$}$] ORC power output.
\item[$\cs{\bm{\lambda}}{$i$}{$\omega,t$}$] ORC interpolation variables.
\item[$\cs{\bm{P}}{bess,d}{$\omega,t$}$] BESS discharge power, on the direct-current side of the converter.
\item[$\cs{\bm{P}}{bess,c}{$\omega,t$}$] BESS charge power, on the direct-current side of the converter..
\item[$\cs{\bm{z}}{bess}{$\omega,t$}$] Charge/discharge indicator for BESS.
\item[$\cs{\bm{E}}{bess}{$\omega,t$}$] BESS stored energy.
\item[$\cs{\bm{P}}{bess,ac}{$\omega,t$}$] BESS power on the AC side of the converter.
\item[$\cs{\bm{a}}{bess}{$\omega,t$}$] BESS operational aging variable.
\item[$\cs{\bm{\Pi}}{bess}{$\omega,t$}$] BESS operational cost.
\item[$\cs{\bm{C}}{bess}{$\omega,t$}$] BESS equivalent carbon emissions.
\item[$\cs{\bm{P}}{pv}{$\omega,t$}$] PV power generation.
\item[$\cs{\bm{P}}{gcp,d-a}{$t$}$] Day-ahead GCP power.
\item[$\cs{\bm{P}}{gcp,imb}{$\omega,t$}$] Power imbalance at the GCP.
\item[$\cs{\bm{P}}{gcp,-}{$\omega,t$}$] Long component of the imbalance power.
\item[$\cs{\bm{P}}{gcp,+}{$\omega,t$}$] Short component of the imbalance power.
\item[$\cs{\bm{z}}{imb}{$\omega,t$}$] Imbalance indicator.
\item[$\cs{\bm{E}}{non-ren}{$\omega,t$}$] Non-renewable energy consumed.
\item[$\cs{\bm{E}}{dc}{$\omega$}$] Daily data center energy consumption.
\end{IEEEdescription}

\section{Introduction}
\subsection{Context and Motivation}
\label{sec:context}
Rapid breakthroughs in \ac{AI} have accelerated the global deployment of data centers. According to the \ac{IEA} \cite{davide_dambrosio_energy_2025}, total data center capacity is projected to grow from approximately \SI{100}{\giga\watt} today to over \SI{225}{\giga\watt} by 2035, while their electricity consumption is expected to grow from \SI{415}{\tera\watt\hour} in 2024 to \SI{1}{\peta\watt\hour} in 2030. This growth is strongly driven by national strategies aimed at gaining competitive advantages in technologies related to \ac{AI}. For example, the European Commission has published an action plan \cite{european_commission_about_nodate} that sets the goal of at least tripling the capacity of data centers across Europe in seven years, supported by an investment initiative of 200 billion Euros. This massive growth is challenging electricity grids, with \acp{TSO} in Ireland, the Netherlands and Germany refusing to connect new large-scale (e.g., $>$ \SI{100}{MW} data centers to the grid \cite{ember_europes_nodate}. Some \acp{TSO} are issuing data center-specific grid codes to regulate connections \cite{ieee_sa_review_2026}. 

Additionally, approximately 90\% of \ac{AI} \ac{WL} is currently training-based, while only about 10\% is inference-based. However, 90\% of \ac{AI} \ac{WL} is expected to become inference-based by 2030 \cite{garg_considerations_2025}. Inference-based \ac{WL} typically requires low-latency processing close to end-users, making small-scale edge data centers (e.g., $<$ \SI{20}{MW}) the preferred solution. These data centers are generally connected to distribution power systems, which are already facing challenges due to volatile distributed generation and fluctuating demand, factors often associated to grid congestion \cite{iea_grid_2025}. Their connection further complicates grid planning and operation for \acp{DSO}.

These trends highlight the need for data centers to operate not only as electricity consumers but also as active/controllable entities. To ensure their scalable deployment, their integration must be supported by strategies that align data center operational flexibility with safe grid operation.
\subsection{Literature Review and Gaps}
\label{sec:LR}
Research on data center integration into electricity markets explores many directions. This review provides an overview of the main topics relevant to the scope of this work.
\subsubsection{Interaction with Electricity Spot Market}
Recent studies have explored how data centers can participate in electricity spot markets through optimized bidding and scheduling. In \cite{chen_multi-objective_2023}, a multi-objective bidding strategy for \ac{WL} scheduling is proposed. Similarly, \cite{zhang_optimal_2020} and \cite{zhang_flexibility_2020} investigate strategies for data center aggregators that leverage geographical \ac{WL} flexibility (i.e., shifting computational tasks across multiple sites) to minimize operational costs. A bidding strategy for data center microgrids with wind and PV generation is presented in \cite{liu_optimal_2024}. However, these approaches are typically not scenario-based and overlook imbalance settlement costs, which are crucial for realistic participation in spot markets.
\subsubsection{\ac{WL} Modeling and Scheduling}
Significant effort has been put on modeling and scheduling job flexibility within data centers. Most studies rely on M/M/n queue models to meet \ac{QoS} objectives \cite{chen_multi-objective_2023, rao_minimizing_2010}, while others require job-level scheduling \cite{liu_renewable_nodate}. The former generally assumes \ac{WL} and hardware homogeneity, whereas the latter becomes computationally intractable for large-scale, day-ahead scheduling \cite{radovanovic_carbon-aware_2023}. The concept of \acp{VCC}, introduced in \cite{radovanovic_carbon-aware_2023} and further exploited in \cite{hall_carbon-aware_2025}, provides a promising trade-off by decoupling day-ahead optimization from real-time \ac{WL} scheduling using virtual capacity limits. In \cite{hall_carbon-aware_2025}, the authors propose a day-ahead planning problem simultaneously designing the \acp{VCC} and the job scheduling strategy while satisfying the constraints in \cite{radovanovic_carbon-aware_2023}. The approach is validated using load profiles from Google clusters, showing \acp{VCC} can be used to efficiently leverage job flexibility in large-scale data centers. These approaches, however, do not fully address the interface with electricity markets (e.g., imbalance settlement costs, coupled operation with local \acp{DER}, bidding strategy, etc.) and focus on large-scale data centers with market access.
\subsubsection{Market Accessibility and Operational Integration}
Several works have examined data centers participation to ancillary services or demand response mechanisms \cite{robert_basmadjian_making_2016, ali_pahlevan_ecogreen_2021, abada_balancing_2024}. These studies generally assume that data centers are knowledgeable market participants. In practice, however, most small-scale data centers lack both expertise and direct market access \cite{takci_data_2025}. Bilateral agreements between \acp{DCO} and grid stakeholders (\acp{DSO}, \acp{TSO}, or energy suppliers/retailers) are thus likely to represent the most realistic entry point for data centers to participate in demand response schemes and/or electricity markets.
\subsubsection{Waste Heat Valorization}
Finally, the sustainable operation of data centers requires valorization of recovered waste heat. Integration with district heating networks \cite{socci_enhancing_2024} and electricity regeneration through \acp{ORC} \cite{wang_coordinated_2023, zakeralhoseini_experimentally_2025} have been studied, with recent \ac{DLC} technologies capable of recovering medium-grade heat further expanding these opportunities \cite{van_erp_co-designing_2020}.
\subsection{Contributions of This Work}
This work addresses several gaps mentioned in \cref{sec:LR}. We propose a day-ahead bidding strategy that coordinates small-scale data center flexibility to enable market participation while accounting for carbon emissions, renewable share and imbalance costs, which are often omitted in similar approaches. A scenario-based, risk-averse \ac{MILP} is formulated to schedule local \ac{BESS}, \ac{PV} generation, and \ac{ORC}, as well as heat exports and flexible \ac{WL}, building on the \acp{VCC} concept from \cite{radovanovic_carbon-aware_2023, hall_carbon-aware_2025} to make day-ahead scheduling computationally tractable. 

Additionally, we propose a bilateral \ac{PPA} between the \ac{DCO} and \ac{DSO}. It grants market access to the \ac{DCO}, while allowing the \ac{DSO} to trigger a virtual de-rating of the transfer capacity at the \ac{GCP}, thus framing a simple demand response mechanism for \ac{DSO}-\ac{DCO} collaboration. The framework is validated through case studies based on the data of a local academic data center, showing advantages for both parties.

\section{Problem Statement}
\subsection{System Description}
\begin{figure}[!t]
    \centering
    \includegraphics[width=\linewidth]{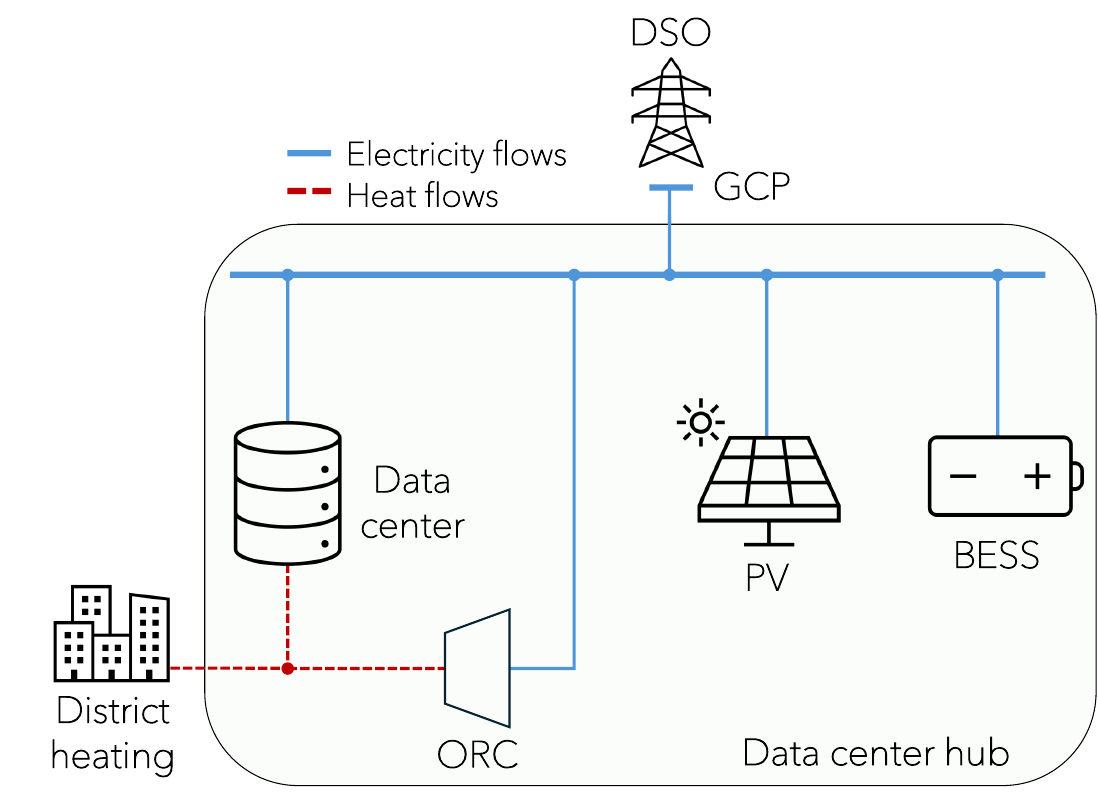}
    \caption{Schematic representation of the system.}
    \label{fig:schematic}
\end{figure}

Given the context highlighted in \ref{sec:context}, the system under study is illustrated in \cref{fig:schematic}. A data center connected to a distribution grid is co-located with local \ac{PV} generation, a local \ac{BESS} and an \ac{ORC}. Waste heat can be recovered from the servers, used by the \ac{ORC} to regenerate electricity and/or sold to the local utility providing district heating services. Given the paper's focus on small-scale data centers connected to the distribution grid (and thus likely neighboring urban areas), the opportunity of recover heat to sell supply local district heating is particularly interesting. This potential can be leveraged by the \ac{DCO} in the \ac{PPA} negotiations.

\subsection{Agreement Between the \ac{DCO} and \ac{DSO}}
\label{sec:agreement}
Small-scale data centers can represent a significant share of the load in power distribution grids. Despite this, they rarely participate in electricity markets and instead rely predominantly on standard \acp{PPA} with fixed pricing structures. While such \acp{PPA} offer contractual simplicity and certainty for \acp{DCO} and \acp{DSO}, their inherent price stability also limits operational incentives. In particular, fixed-price \acp{PPA} do not encourage improvements in power consumption predictability or the provision of operational flexibility. As a result, the potential of small-scale data centers to act as active and dispatchable grid participants remains largely unexploited. We propose and study a custom \ac{PPA} between the \ac{DCO} and \ac{DSO}, with the following features.
\begin{itemize}
\item \textbf{\ac{DCO} market access:}
Under the proposed agreement, the data center is granted access to the day-ahead electricity market. Given its size (i.e., $<$ \SI{20}{\mega\watt}), the data center operates as a price-taker in the spot market. It performs a day-ahead scheduling of resources within the data center hub and submits a corresponding power profile bid to the \ac{DSO}, which acts as an interface with the market platform. 
Following market clearing, the \ac{DCO} is financially responsible for deviations from the cleared day-ahead schedule. Imbalances are therefore settled according to imbalance prices, incentivizing accurate load forecasting and schedule tracking.

\item \textbf{Dynamic rating of the \ac{GCP}:}
In exchange for market access, the \ac{DSO} retains the right to apply a day-ahead virtual de-rating of the data center \ac{GCP}, reducing the maximum import capacity during selected hours. The de-rating signal is communicated day-ahead and must be incorporated into the \ac{DCO}'s bidding process.

To ensure contractual fairness and operational predictability, the de-rating profile is bounded by a minimum capacity guarantee and cumulative de-rating daily and weekly budgets. Defining the de-rating amount in \sref{eq:curtailment-amt}, we formalize three contractual limits \sref{eq:min rating}-\sref{eq:weekly-de-rating}. 

\begin{align}
\Delta \cs{P}{gcp}{$t$
}&=\cs{P}{gcp}{rated}-\cs{P}{gcp,cap}{$t$} \label{eq:curtailment-amt}\\
\cs{P}{gcp,cap}{$t$} 
&\ge \cs{P}{gcp,min}{},~ \forall t \label{eq:min rating}\\
\sum_{t\in\mathcal{D}} \Delta \cs{P}{gcp}{$t$}\Delta T
&\le (\cs{P}{gcp}{rated}-\cs{P}{gcp,min}{}) \cs{t}{daily,lim}{} \label{eq:daily-de-rating}\\
\sum_{t\in\mathcal{W}} \Delta \cs{P}{gcp}{$t$}\Delta T
&\le (\cs{P}{gcp}{rated}-\cs{P}{gcp,min}{}) \cs{t}{weekly,lim}{} \label{eq:weekly-de-rating}
\end{align}

A minimum import capacity is guaranteed in \sref{eq:min rating}, while \sref{eq:daily-de-rating} and \sref{eq:weekly-de-rating} limit the cumulative daily and weekly de-rating that can be imposed by the \ac{DSO}. 
\end{itemize}
In \cref{sec:method}, we propose a stochastic optimization-based bidding strategy for data center hubs with such \ac{PPA}. In \cref{sec:study-cases}, we study the potential of such agreement using real data from an academic data center at EPFL in Switzerland, and show that it can benefit both \acp{DCO} and \acp{DSO}.

\section{Day-Ahead Bidding Strategy}
\label{sec:method}
\subsection{Objective}
The objective in \sref{eq:obj} minimizes the expected daily operational costs across scenarios $\omega \in \Omega$, while incorporating risk-aversion through the $\alpha$-level \ac{CVaR}\cite{rockafellar_optimization_2000}. The relative importance of risk-aversion can be set by the user through the parameter $\beta \in [0, 1]$\footnote{Note that $\beta$ can be tuned via pareto front studies. We do not show such studies in this paper because of page limits.}. We highlight decision variables in bold. 
\begin{equation}
\label{eq:obj}
\min \quad (1-\beta) \, \mathbb{E}(\bm{\Pi}_\text{op}^{\omega}) + \beta \, \text{CVaR}_\alpha(\bm{\Pi}_\text{op}^{\omega})
\end{equation}

Daily operational costs $\bm{\Pi}_\text{op}^{\omega}$ are defined as the sum over $t\in\mathcal{T}$ of $\bm{\Pi}_\text{op}^{\omega, t}$, defined in \sref{eq:opex}. They include electricity costs (with day-ahead and imbalance components), energy resources aging costs, operational carbon emissions costs (with grid imports and energy resources aging components) and revenues generated by selling waste heat. 
\begin{equation}
\label{eq:opex}
\begin{aligned}
    \bm{\Pi}_\text{op}^{\omega, t} =& \bm{\Pi}_\text{d-a}^{\omega,t} + \bm{\Pi}_\text{imb}^{\omega,t} + \cs{\bm{\Pi}}{op, DERs}{$\omega,t$} - \bm{\Pi}_\text{heat}^{\omega,t}\\
    +& \pi_\text{carbon}\Big(\bm{C}_\text{gcp}^{\omega,t} + \cs{\bm{C}}{op, DERs}{$\omega,t$} \Big)
\end{aligned}
\end{equation}
Note that $\pi_\text{carbon}$ represents the cost of carbon emissions, expressed in \si{\univcurr\per{\gram\COtwoeq}}, where \si{\univcurr} is the universal currency symbol.
\subsection{Data Center Model}
\subsubsection{Computational Resources}
We define the data center as a set of clusters $\mathcal{C}$. Each cluster c has its own configuration and resources, with capacity $\kappa^{\text{c, res}}$. Clusters can contain \acp{CPU} and \acp{GPU}. Memory resources are split in two categories, $\mathcal{R}_\text{mem}=\{\text{MEM-CPU, MEM-GPU}\}$, as \acp{CPU} and \acp{GPU} typically use different types of memory. We define the set of computational resources types $\mathcal{R}=\{\text{CPU, GPU}\}$. Within clusters, resources of a given nature should be homogeneous (e.g., a single \ac{CPU} model per cluster). The usage of resources per cluster $\cs{\bm{u}}{}{$\omega,t$,c,res}$ is the decision variable in the data center model. 
\subsubsection{Workload}
Following \cite{radovanovic_carbon-aware_2023}, we model the data center \ac{WL} by defining two broad categories: (i) inelastic \ac{WL} (i.e., jobs that must be executed upon arrival and, therefore, offer no scheduling flexibility) and (ii) flexible \ac{WL} (i.e., jobs that must be completed within a given time window, allowing flexibility regarding their start time). The inelastic \ac{WL}’s per cluster resource demand $I^{\omega,t,\text{c,res}}$ is assumed to be forecastable as a time series, 24h-ahead. For flexible \ac{WL}, we focus  on \ac{WL} that has to be completed within a day. We assume that the daily cumulative demand of computational resources $F^{\omega,\text{c,res}}$ can be forecasted 24h-ahead. These assumptions are supported by the work conducted in \cite{radovanovic_carbon-aware_2023}. 

\begin{subequations}
\begin{align}
& \cs{\bm{u}}{}{$\omega,t$, c, res} \geq I^{\omega,t,\text{c,res}}\label{const:inelastic-wl} \\
& \sum_{t\in\mathcal{T}} \Big(\cs{\bm{u}}{}{$\omega,t$, c, res} - I^{\omega,t,\text{c,res}}\Big)  \geq \frac{F^{\omega,\text{c,res}}}{\Delta T} \label{const:flexible-wl} \\
& \sum_{t\in\mathcal{T}} \Big(\cs{\bm{u}}{}{$\omega,t$, c, res} - I^{\omega,t,\text{c,res}}\Big)  \leq \frac{F^{\omega,\text{c,res}}}{\Delta T} \label{const:overprovision}
\end{align}
\end{subequations}

When verified for all $\omega \in \Omega,~t \in \mathcal{T},~\text{c} \in \mathcal{C}, ~\text{res} \in \mathcal{R}$, \sref{const:inelastic-wl} and \sref{const:flexible-wl} ensure the satisfaction of the two \ac{WL} categories strictly, while \sref{const:overprovision} avoids over-using resources (and, in this case, \sref{const:flexible-wl} and \sref{const:overprovision} could be joined in a single equality constraint). However, to embed \ac{QoS} considerations, we replace \sref{const:inelastic-wl} and \sref{const:flexible-wl} by the \acp{CC} \sref{const:inelastic-wl-cc} and \sref{const:flexible-wl-cc} (i.e., we relax \ref{const:inelastic-wl} and \ref{const:flexible-wl}). Practically, both \acp{CC} are implemented using violation indicator variables and ensuring the likelihood of violations does not exceed the allowed quotas.
\begin{subequations}
    \begin{align}
        &\mathbb{P}(\ref{const:inelastic-wl}) \geq \Gamma_\text{inelastic} \label{const:inelastic-wl-cc}\\
        &\mathbb{P}(\ref{const:flexible-wl}) \geq \Gamma_\text{flexible} \label{const:flexible-wl-cc}
    \end{align}
\end{subequations}
The usage of memory resources is tracked and modeled using a linear mapping \sref{const:mem-follow-res} to the related resource (e.g., MEM-CPU usage is proportional to CPU usage) using historical average ratios $\cs{\gamma}{mem}{c, res}$. 
\begin{equation}
    \begin{aligned}
        \cs{\bm{u}}{}{$\omega,t$, c, res} =& \cs{\gamma}{mem}{c, res}\cs{\bm{u}}{}{$\omega,t$, c, rel-res}, \\ & \forall \omega \in \Omega, t\in\mathcal{T}, \text{c}\in \mathcal{C}, \text{res} \in \mathcal{R}_\text{mem} \label{const:mem-follow-res}
    \end{aligned}
\end{equation}

The usage of resources must be less or equal to the available capacity for each resource and cluster, leading to \sref{const:res-available-capacity}.
\begin{equation}
   \cs{\bm{u}}{}{$\omega,t$, c, res} \leq \kappa^{\text{c, res}}, \forall \omega \in \Omega,t \in \mathcal{T},~\text{c}\in\mathcal{C},~\text{res} \in \mathcal{R} \label{const:res-available-capacity}
\end{equation}
\subsubsection{Power Consumption}
The power consumption of the data center is modeled as a function of resource usage $\cs{\bm{u}}{}{$\omega,t$, c, res}$. In this work, we assume an affine relation \sref{const:dc-power-cons} between resource usage and cluster IT-equipment power consumption \cite{radovanovic_carbon-aware_2023, dayarathna_data_2016}. Note that $\eta_\text{PUE}$ is the \ac{PUE} of the data center and includes the consumption of non-IT equipment (e.g., fans, pumps, etc). In what follows, we omit the explicit quantifiers $\forall \omega \in \Omega,t \in \mathcal{T}$ for compactness. All constraints (with the exception of \acp{CC}) involving scenario- and time-dependent variables hold for all $\omega \in \Omega$ and $t \in \mathcal{T}$.
\begin{equation}
\begin{aligned}
\bm{P}_\text{dc}^{\omega,t} =& \eta_\text{PUE}
\sum_{\text{c}\in\mathcal{C}}
\big(\rho_{\text{inter}}^\text{c} + \sum_{\text{res}\in\mathcal{R} \cup \mathcal{R}_\text{mem}}\rho_{\text{coeff}}^{\text{c, res}}
\cs{\bm{u}}{}{$\omega,t$, c, res}\big)
\label{const:dc-power-cons}
\end{aligned}
\end{equation}
\subsubsection{Heat Recovery}
We consider the case of data centers equipped with \ac{DLC} for computational resources, as this technology is more efficient and allows for the recovery of medium-grade heat, which can in turn be exploited by the \acp{ORC} to regenerate electricity (see \cref{fig:schematic}). Therefore, reusable heat comes from the power losses of the resources that are cooled using this technology (e.g., \acp{CPU} and \acp{GPU}). We define the recovered heat in \sref{const:heat-recovery}. Note that, in practice, the idle consumption of cooled resources in each cluster $\rho_\text{rec,idle}^\text{c}$ can be estimated empirically by measuring the heat recovered with the cluster operating in idle mode.
\begin{equation} 
\begin{aligned}
    \cs{\bm{\dot Q}}{rec}{$\omega,t$}=&\sum_{\text{c}\in\mathcal{C}} \eta_\text{rec}^\text{c, res} \left(\rho_\text{rec,idle}^\text{c}+\sum_{\text{res}\in\mathcal{R}_\text{cooled}}\rho_{\text{coeff}}^{\text{c, res}}\cs{\bm{u}}{}{$\omega,t$, c, res}\right) \label{const:heat-recovery}
\end{aligned}
\end{equation}

Recovered heat can be (i) used by the ORC, (ii) sold to local utilities operating district heating or (iii) dumped and lost \sref{const:heat-usage}.
\begin{equation}
\begin{aligned}
    \label{const:heat-usage}
    \cs{\bm{\dot Q}}{rec}{$\omega,t$} =& \cs{\bm{\dot Q}}{orc, in}{$\omega,t$}+ \cs{\bm{\dot Q}}{sold}{$\omega,t$} + \cs{\bm{\dot Q}}{lost}{$\omega,t$}
\end{aligned}
\end{equation}

Note that, in practice, residual heat at the output of the \ac{ORC} could be used for district heating purposes if its temperature is high enough. In this formulation, we assume the residual heat of the \ac{ORC} cannot be used by the district heating.
\subsection{Energy Resources Models}
\subsubsection{\ac{ORC} Model} 
The \ac{ORC} uses recovered heat to generate electrical power. The relation between the two, if the \ac{ORC} is operated at maximum efficiency, is non-linear \cite{muhammad_design_2015}. We thus sample the curve $P_\text{orc}(\dot Q_\text{orc, in})$ and get the sample set $\{(\dot q_i, p_i): i\in[0, 1, ..., N_s],~p_i=P_\text{orc}(\dot q_i)\}$. The samples are then used to formulate \sref{const:orc-piecewise}, where \sref{const:orc-interp} constrains the interpolation variables $\bm{\lambda}_i^{\omega, t}$, \sref{const:orc-heat} interpolates the input heat and \sref{const:orc-power} defines the upper bound for the output power. While not made explicit here, \ac{SOS2} piecewise constraints are used to ensure that two consecutive $\bm{\lambda}_i^{\omega,t}$ at maximum can be non-zero simultaneously \cite{beale_global_1976, bynum_pyomo_2021}. 
\begin{subequations}
\label{const:orc-piecewise}
\begin{align}
    & \sum_{i=0}^{N} \bm{\lambda}_i^{\omega,t} = 1, \quad \bm{\lambda}_i^{\omega,t} \geq 0, \label{const:orc-interp}\\
    & \bm{\dot Q}_\text{orc, in}^{\omega,t} = \sum_{i=0}^{N} \dot q_i \, \bm{\lambda}_i^{\omega,t}, \label{const:orc-heat}\\
    & \bm{P}_\text{orc}^{\omega,t} \leq \sum_{i=0}^{N} p_i \, \bm{\lambda}_i^{\omega,t} \label{const:orc-power}
\end{align}
\end{subequations}

We highlight that we only define an upper bound for the output power, implying that, for a given heat input, the \ac{ORC} can be controlled to track any setpoint between zero and the upper bound. 

\subsubsection{\ac{BESS} Model}
We model the energy dynamics of the \ac{BESS} using a bucket model in \sref{const:bess-dynamics}. The \ac{BESS} state-of-energy and power limits are ensured by \sref{const:bess-energy-bounds} and \sref{const:bess-power-bounds}, respectively. Finally, we account for the \ac{BESS} one-way efficiency $\eta_\text{bess}$ in \sref{const:bess-ac-dc}, where $\cs{\bm{P}}{bess, d}{$\omega,t$}$ and $\cs{\bm{P}}{bess, c}{$\omega,t$}$ denote the discharge and charge power of the battery. We guarantee their mutual exclusivity through binary activation constraints via Big-M formulation \sref{const:bess-charge-discharge}--\sref{const:big-m-bess-2}, where $\cs{\bm{z}}{bess}{$\omega,t$}$ is a binary indicator variable (here, $\cs{P}{bess}{rated}$ plays the role of the Big-M constant since \ac{BESS} power is constrained by its rated value). Finally, we ensure, through \sref{const:bess-avg-zero} along with the inherent inclusion of \ac{BESS} losses via \sref{const:bess-ac-dc}, that the \ac{BESS} does not get charged or discharged on average. This ensures that the \ac{BESS}, in expectation, remains a neutral energy buffer, guaranteeing its availability for subsequent daily operation.
\begin{subequations}
    \begin{align}
        &\cs{\bm{E}}{bess}{$\omega,t+\Delta T$} =\cs{\bm{E}}{bess}{$\omega,t$} + \Delta T \cs{\bm{P}}{bess}{$\omega,t$}, \label{const:bess-dynamics}\\
        &\cs{E}{bess}{min} \leq\cs{\bm{E}}{bess}{$\omega,t$} \leq \cs{E}{bess}{max}, \label{const:bess-energy-bounds}\\
        -&\cs{P}{bess}{rated}\leq\cs{\bm{P}}{bess}{$\omega,t$}\leq \cs{P}{bess}{rated}, \label{const:bess-power-bounds}\\
        &\cs{\bm{P}}{bess, ac}{$\omega,t$} =\eta_\text{bess}^{-1}\cs{\bm{P}}{bess, c}{$\omega,t$}-\eta_\text{bess}\cs{\bm{P}}{bess, d}{$\omega,t$}, \label{const:bess-ac-dc}\\
        &\cs{\bm{P}}{bess}{$\omega,t$} =\cs{\bm{P}}{bess, c}{$\omega,t$} - \cs{\bm{P}}{bess, d}{$\omega,t$} \label{const:bess-charge-discharge}\\
        &\cs{\bm{P}}{bess, c}{$\omega,t$}\leq \cs{\bm{z}}{bess}{$\omega,t$}\cs{P}{bess}{rated}\label{const:big-m-bess-1}\\
        &\cs{\bm{P}}{bess, d}{$\omega,t$}\leq (1-\cs{\bm{z}}{bess}{$\omega,t$})\cs{P}{bess}{rated}\label{const:big-m-bess-2}\\
        &\mathbb{E}\Big(\sum_{t\in\mathcal{T}}\bm{P}_\text{bess}^{\omega,t}\Big) = 0 \label{const:bess-avg-zero} 
    \end{align}
\end{subequations}

We model the aging of the \ac{BESS} due to cycling, as it corresponds to the operational aging of the asset. To do so, we assume the aging of the asset to be linear with the system's energy throughput. This is made explicit in \sref{const:bess-aging}, where $\cs{L}{bess}{cycles}$ and $\cs{E}{bess}{rated}$ are the rated number of cycles and the rated capacity of the asset, respectively. Then, the operational cost of the battery $\bm{\Pi}_\text{bess}^{\omega,t}$ and its operational carbon footprint $\bm{C}_\text{bess}^{\omega,t}$ (scope 3 emissions) are computed in \sref{const:bess-opex} and \sref{const:bess-carbon}. Note that $\bm{\Pi}_\text{bess}^{\omega,t}$ and $\bm{C}_\text{bess}^{\omega,t}$ are part of the total operational costs and emissions of the data center hub \acp{DER} in the objective \sref{eq:obj}.
\begin{subequations}
    \begin{align}
        &\cs{\bm{a}}{bess}{$\omega,t$} = \frac{|\cs{\bm{P}}{bess}{$\omega,t$}|\Delta T}{2 \cs{L}{bess}{cycles} \cs{E}{bess}{rated}}, \label{const:bess-aging} \\
        &\bm{\Pi}_\text{bess}^{\omega,t} = \cs{\bm{a}}{bess}{$\omega,t$} \Pi_\text{bess}^\text{investment}, \label{const:bess-opex}\\
        &\bm{C}_\text{bess}^{\omega,t} = \cs{\bm{a}}{bess}{$\omega,t$} C_\text{bess}^\text{LCA} \label{const:bess-carbon}
    \end{align}
\end{subequations}
\subsubsection{\ac{PV} Generation Model}
We model \ac{PV} generation using the simplified linear \ac{PV} performance model described in \sref{const:pv-power}, and PV generation can be curtailed. 
\begin{equation}
    \cs{\bm{P}}{pv}{$\omega,t$} \leq \frac{\cs{i}{ghi}{$\omega,t$}}{\cs{i}{ghi, ref}{}}\cs{P}{pv}{rated} \label{const:pv-power}
\end{equation}

\subsection{Electricity Supply Model}
\subsubsection{Electricity Market}
We model the electricity spot market as well as the imbalance costs faced by the market player (i.e., the \ac{DCO}), which is therefore modeled as a passive player in the balancing markets. We base our model on the european (and swiss) electricity markets. In \sref{const:gcp-power-split}, we decompose the power at the \ac{GCP} \sref{const:sum-of-res-power} into a day-ahead scenario-independent power component $\cs{\bm{P}}{gcp, d-a}{$t$}$ and an imbalance power component $\cs{\bm{P}}{gcp, imb}{$\omega,t$}$. The day-ahead power component is billed at the spot price $\pi_\text{spot}^{\omega,t}$ in \sref{const:spot-opex}. In \sref{const:gcp-long-short}, the imbalance power is decomposed into a long $\cs{\bm{P}}{gcp,-}{$\omega,t$}$ and short $\cs{\bm{P}}{gcp,+}{$\omega,t$}$ component to match the settlement structure of imbalance costs, detailed in \sref{const:imb-opex}. The mutual exclusivity of $\cs{\bm{P}}{gcp,-}{$\omega,t$}$, $\cs{\bm{P}}{gcp,+}{$\omega,t$}$ is guaranteed through \sref{const:big-m-gcp-1}--\sref{const:big-m-gcp-2}, where $\cs{\bm{z}}{imb}{$\omega,t$}$ is a binary indicator variable and $\cs{P}{gcp}{rated}$ is the rated power of the \ac{GCP}. Finally, \sref{const:dso-cap} constrains the \ac{GCP} power using dynamic virtual transfer capacity values $\cs{P}{gcp, cap}{$t$}$ set by the grid operator, according to the bilateral agreement in \cref{sec:agreement}.
\begin{subequations}
    \begin{align}
        &\cs{\bm{P}}{gcp}{$\omega,t$} = \cs{\bm{P}}{dc}{$\omega,t$}-\cs{\bm{P}}{orc}{$\omega,t$}+\cs{\bm{P}}{bess, ac}{$\omega,t$}-\cs{\bm{P}}{pv}{$\omega,t$}\label{const:sum-of-res-power}\\
        &\cs{\bm{P}}{gcp}{$\omega,t$} = \cs{\bm{P}}{gcp, d-a}{$t$} + \cs{\bm{P}}{gcp, imb}{$\omega,t$}, \label{const:gcp-power-split}\\
        &\cs{\bm{P}}{gcp, imb}{$\omega,t$} = \cs{\bm{P}}{gcp,+}{$\omega,t$} - \cs{\bm{P}}{gcp,-}{$\omega,t$}, \label{const:gcp-long-short}\\
        &\bm{\Pi}_\text{d-a}^{\omega,t} = \pi_\text{spot}^{\omega,t}\cs{\bm{P}}{gcp, d-a}{$t$}\Delta T, \label{const:spot-opex}\\
        &\bm{\Pi}_\text{imb}^{\omega,t} = \Delta T(\pi_\text{short}^{\omega,t}\cs{\bm{P}}{gcp,+}{$\omega,t$} - \pi_\text{long}^{\omega,t}\cs{\bm{P}}{gcp,-}{$\omega,t$}) \label{const:imb-opex}\\
        &\cs{\bm{P}}{gcp,+}{$\omega,t$}\leq\cs{\bm{z}}{imb}{$\omega,t$}\cs{P}{gcp}{rated}\label{const:big-m-gcp-1}\\
        &\cs{\bm{P}}{gcp,-}{$\omega,t$}\leq(1-\cs{\bm{z}}{imb}{$\omega,t$})\cs{P}{gcp}{rated}\label{const:big-m-gcp-2}\\
        &\cs{\bm{P}}{gcp}{$\omega,t$}\leq \cs{P}{gcp, cap}{$t$}\label{const:dso-cap}
    \end{align}
\end{subequations}
\subsubsection{Carbon Emissions}
We compute the scope 2 carbon emissions due to electricity imports in \sref{const:gcp-carbon}.
\begin{equation}
        \cs{\bm{C}}{gcp}{$\omega,t$}= c_\text{gcp}^{\omega,t}\cs{\bm{P}}{gcp}{$\omega,t$} \Delta T \label{const:gcp-carbon}
\end{equation}

\subsection{Heat Market Model}
We assume that the data center is connected to a district heating network operated by the local utility provider. The data center and the utility provider have a simple fixed unit price contract, allowing the data center to sell medium-grade heat. The revenue stream from sold heat is computed in \sref{const:heat-revenue}. The amount of heat that can be sold is limited by the district heating demand \sref{const:max-heat-demand}.
\begin{subequations}
\begin{align}
    &\bm{\Pi}_\text{heat}^{\omega,t} = \pi_\text{heat}\bm{\dot Q}_\text{sold}^{\omega,t}\Delta T \label{const:heat-revenue}\\
    &\cs{\bm{\dot Q}}{sold}{$\omega,t$} \leq \cs{\dot Q}{demand}{$\omega, t$} \label{const:max-heat-demand}
\end{align}
\end{subequations}

\subsection{Renewable Share}
We track the renewable share of the data center energy consumption. Since local generation is fully renewable (from the \ac{PV} and the \ac{ORC}), we track the non-renewable electricity $\cs{\bm{E}}{non-ren}{$\omega,t$}$ that flows into the system from the \ac{GCP} \sref{const:non-renewable-energy}. Note that $0\leq \cs{s}{gcp}{$\omega,t$}\leq 1$ is the renewable share of the electricity imported from the grid. We formulate a \ac{CC} on the non-renewable energy consumption \sref{const:non-ren-cc}, where $\cs{\bm{E}}{non-ren}{$\omega$}$ is the non-renewable energy at the end of the day and $\cs{\bm{E}}{dc}{$\omega$}$ is the total daily energy consumption of the data center. This constraint ensures that the renewable share cannot be smaller than the set target $S_\text{cap}$ in more than $100\alpha_r$\si{\percent} of the scenarios, where $\alpha_r$ denotes the allowed fraction of violations. Note that \sref{const:non-ren-cc} can induce infeasibility and, therefore, $S_\text{cap}$ must be selected with care. Practically, we implement the corresponding \ac{CC} by introducing violation indicators and ensuring that the probability of a violation occuring is below $\alpha_r$.
\begin{subequations}
\begin{align}
    &\cs{\bm{E}}{non-ren}{$\omega, t+1$} = \cs{\bm{E}}{non-ren}{$\omega,t$}+(1-\cs{s}{gcp}{$\omega,t$})\cs{P}{gcp}{$\omega,t$}\Delta T \label{const:non-renewable-energy}\\
    &\mathbb{\bm{P}}(\cs{\bm{E}}{non-ren}{$\omega$}\geq (1-\cs{S}{cap}{})\cs{\bm{E}}{dc}{$\omega$}) \leq \alpha_r \label{const:non-ren-cc}
\end{align}
\end{subequations} 
\subsection{Final Problem Formulation}
We formulate the full optimization problem \sref{prob:dispatch}.
\begin{equation}
\label{prob:dispatch}
\begin{aligned}
\min \quad & \sref{eq:obj} \\
\text{s.t.} \quad
& \sref{eq:opex}\\
& \sref{const:overprovision}
  \text{ to } \sref{const:non-ren-cc}
\end{aligned}
\end{equation}
Note that, once the day-ahead bidding process is performed, we can construct \acp{VCC} for the \acp{CPU} and \acp{GPU} of each cluster. In \sref{eq:vcc}, we propose a conservative construction, where, for each timestep, the \ac{VCC} is set to the maximum resource usage across all scenarios. We note that, in practice, the \acp{VCC} are derived after solving the proposed method and their precise construction should be adapted to the performance of the real-time control layer.
\begin{equation}
    \begin{aligned}
        \cs{\kappa}{v}{$t$, c, res} =& \text{max}_{\omega\in\Omega}(\cs{\bm{u}}{}{$\omega,t$, c, res}), \\ &\forall t\in\mathcal{T},~\text{c}\in\mathcal{C}, ~\text{res}\in\mathcal{R}\label{eq:vcc}
    \end{aligned}
\end{equation}

\section{Study Cases}
\label{sec:study-cases}
We present the case of an academic \SI{200}{\kilo\watt}-rated data center located at EPFL, in Lausanne, Switzerland. The grid connection power rating is \SI{300}{\kilo\watt}. An \ac{ORC} with a rated maximum input heat flow of \SI{100}{\kilo\watt}, \SI{200}{\kilo\watt} of \ac{PV} generation and a \ac{BESS} with a \SI{250}{\kilo\watt} rated power and \SI{250}{\kilo\watt\hour} capacity are co-located with the data center. The data center, the \ac{ORC}, the \ac{PV} and the \ac{BESS} are controllable. 

The simulations ran on a MacBook Pro with Apple M2 Max Chip, 32 GB of RAM, using the GUROBI solver.
\subsection{Data Availability}
Unless specified otherwise, hourly data from January 2023 to August 2025 is used.
\subsubsection{Electricity Imports}
We recover the electricity spot prices and imbalance prices for Switzerland from the ENTSO-E transparency platform \cite{hirth_entso-e_2018}. We recover the day-ahead forecasts of renewable generation for Germany, Italy, France and Austria, as candidate explanatory variables for the spot price. We use \cite{emissium_emissium_nodate} to get the dynamic carbon intensity (GWP100a) and renewable share of electricity in canton Vaud, Switzerland.
\subsubsection{Carbon Cost}
\label{sec:carbon-param}
The cost of a gram of \si{\COtwoeq} emitted ($\pi_\text{carbon}$) can be set in various ways\footnote{For example, the EU \ac{ETS} sets prices via traded allowances, while voluntary carbon markets offer credits for avoidance or removal (e.g., direct air capture or reforestation), with prices varying from hundreds to thousands of Euros per ton of \si{}\COtwoeq.}. In this paper, we use $\pi_\text{carbon}$ = \SI{265}{EUR/t\COtwoeq}, the mean social cost of carbon from \cite{rennert_comprehensive_2022}.
\subsubsection{Heat Price}
\label{sec:heat-price}
According to \cite{ewb_tarife_2024}, district heating heat is sold at \SI{0.15}{EUR/kWh} to end-consumers. We assume the data center has secured a deal to sell its recovered heat to the district heating operator at \SI{0.03}{EUR/kWh}.
\subsection{Scenario Generation}
While forecasting is not the focus of this work, its performance impacts the results of the stochastic optimization method. We thus describe the methods we use for the sake of reproducibility.
\subsubsection{Solar Irradiance}
We use the service from \cite{paxian_dwd_2023} to recover up to 40 day-ahead irradiance scenarios.
\subsubsection{Spot Prices, Renewable Share and Carbon Intensity}
\label{subsec:gbr}
We implement a \ac{GBR} forecasting method \cite{ke_lightgbm_nodate} to forecast the spot prices, the renewable share and the carbon intensity of the electricity imported by the data center under study. We follow the following steps, for each of the stochastic variables, independently.
\begin{itemize}
    \item We augment the data set adding candidate features, such as day-ahead renewable forecasts from neighboring countries, lagged data of the stochastic variable (week-ahead lag and day-ahead lag) and day of the week. We standardize the data to zero means and unit variances.
    \item We split the data into a training and a test data set, with the day to plan as the first test date.
    \item Using the training data set, we identify the five dominant features by running a first \ac{GBR} with all candidate features. We drop all the other candidates. 
    \item We run \ac{GBR} using the selected features only. We store the non-standardized residuals of the training set.
    \item The expected value of the stochastic variable is predicted for the target day and rescaled to its original units.
    \item We build $N$ scenarios by bootstrap subsampling $N$ days in the set of residuals. This preserves the realistic time-coupling of residuals in the scenarios.
\end{itemize}
\subsubsection{Imbalance Prices}
Long and short imbalance price data are collected for the period from January 2023 to August 2025 \cite{hirth_entso-e_2018}. Given the high volatility typically observed in imbalance prices, we adopt a simplified modeling approach in which imbalance costs are assumed to be proportional to the spot price. The proportionality factors are calibrated such that, when the approximation is applied to the historical data set, 40\% of the observations result in an underestimation of the actual imbalance prices.
\subsubsection{Data Center \ac{WL}}
As anticipated earlier, we use data from a data center providing its services to the research community at EPFL\footnote{https://www.epfl.ch/research/facilities/rcp/}. We recover node-level statistics for power consumption, CPU usage, GPU usage, CPU memory used, GPU memory used. The node configuration determines to which cluster a given node belongs. The capacity of the clusters is detailed in \cref{tab:dc-config}. The data have a granularity of \SI{2}{\min} and span from October 16th 2024 to August 4th 2025. 

\begin{table}[ht]
\caption{Cluster Capacities of the Data Center Under Study}
\centering
\begin{tabular}{l|l|l|l|l|l}
Cluster & Nodes & GPU & CPU cores & GPU-MEM & CPU-MEM \\ \hline
A100       & 32       & 256          & 1024               & 10240            & 32768            \\ \hline
H100       & 10       & 80           & 960                & 6400             & 15360            \\ \hline
V100       & 16       & 64           & 576                & 2048             & 6160             \\ \hline
\end{tabular}
\label{tab:dc-config}
\end{table}
Using the nodes configurations, we aggregate resources into clusters and downsample the data to an hourly resolution. We then fit three linear models to map resource usage to cluster-level power consumption, determining the coefficients in \sref{const:dc-power-cons}. In addition, we compute the average ratio of memory to computational resource usage $\cs{\gamma}{mem}{res, c}$, which is required in \sref{const:mem-follow-res}. Finally, we use the \ac{GBR} method (cf. \ref{subsec:gbr}) to generate scenarios for hourly inelastic \ac{WL} resources request and daily flexible \ac{WL} resources request.
\subsubsection{Scenario Combination and Clustering} 
To be consistent with \cite{paxian_dwd_2023}, we generate 40 scenarios for each stochastic parameter (i.e., spot prices, renewable share, carbon intensity, solar irradiance, inelastic CPU/GPU demand and flexible CPU/GPU demand). We generate 500'000 combinations of these scenarios (assuming the parameters are independent) and cluster them using K-Means with 60 clusters. Then, for each cluster, we select the closest scenario (using the $L^2-$norm) to the centroid and assign its probability according to the cluster population. 
\subsection{Study Cases Discussion}
We evaluate the proposed bidding strategy and \ac{PPA} across three case studies. The first one provides a statistical analysis of the performance of the bidding strategy with different \acp{PPA} over the month of July 2025. The second aims at quantifying the value of flexible \ac{WL} in the day-ahead electricity market. The final case study analyzes the impact of virtual de-rating on day-ahead planning. In the first two case studies, ex-ante and ex-post operational metrics are compared. Ex-post performance is assessed by enforcing the day-ahead bid determined in the stochastic optimization planning stage and re-optimizing operational decisions using the realized values of \ac{GHI}, spot prices, carbon intensity, renewable share and \ac{WL}.

\subsubsection{Study Case I --- on the Advantages of the Proposed \ac{PPA}}
\label{study-case:market-v-tou}
In this study case, we evaluate the contractual framework proposed in \cref{sec:agreement} from the perspective of the data center. We compare key operational metrics for two \acp{PPA}: (i) the proposed \ac{PPA} and (ii) \ac{ToU} fixed tariffs representative of 2025 \cite{romande_energie_tarifs_nodate}. For the sake of simplicity, the analysis is restricted to energy supply, transmission/imbalance costs and \ac{BESS} aging costs. For (i), the data center pays the market-cleared spot prices and the settled imbalance costs, while for (ii) it faces fixed \ac{ToU} tariffs. Importantly, while the supply agreements are different, both cases leverage the method proposed in this paper. We run the method for 23 days in July 2025. To isolate the impact of the agreements, no virtual de-rating request from the \ac{DSO} is considered (thereby estimating the maximum gains that the data center can achieve). We consider 50\% of flexible \ac{WL} in this study case and will use these results as a baseline for the study case in \cref{study-case:flex-value}. 

\begin{figure}[ht]
    \centering
    \includegraphics[width=0.8\linewidth]{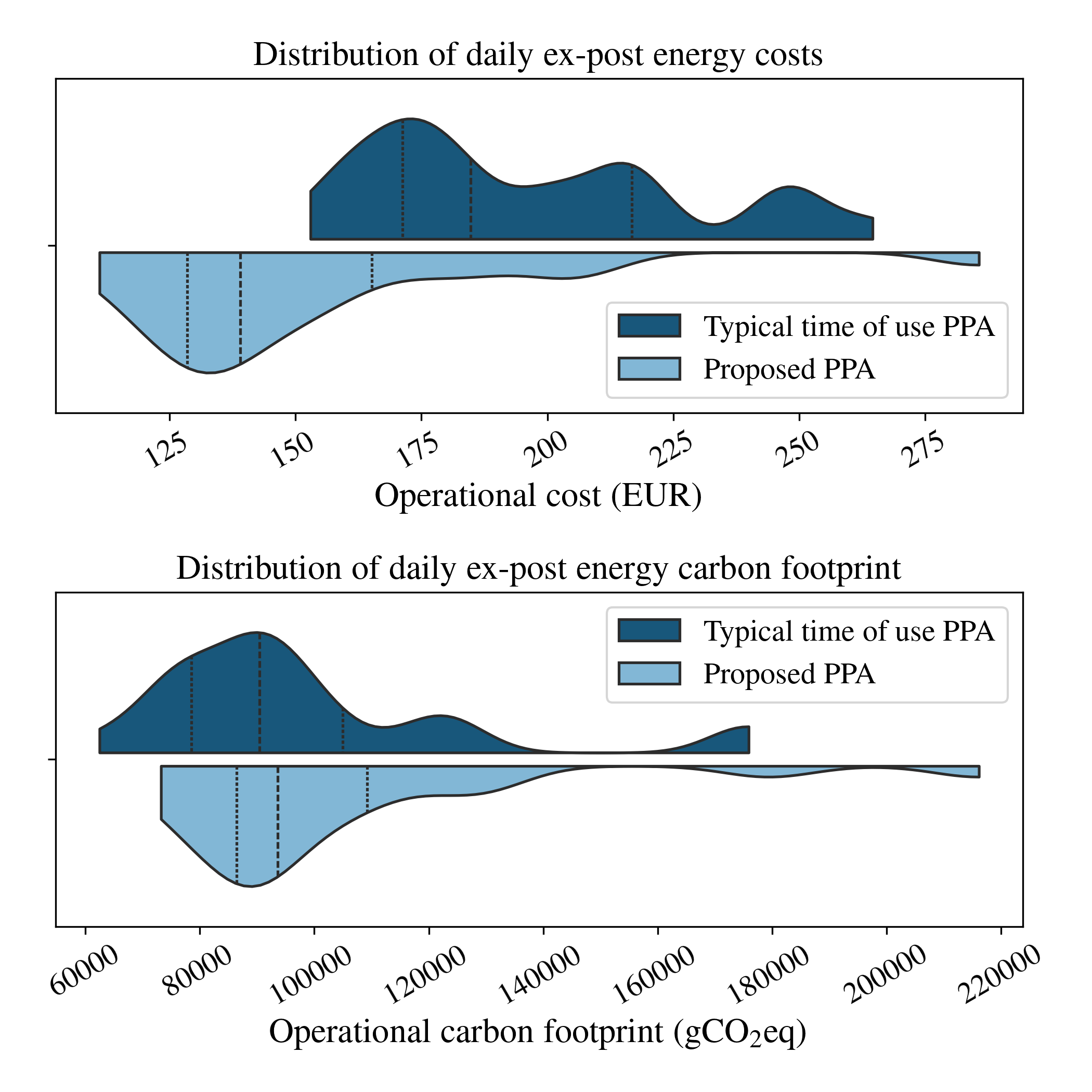}
    \caption{Distribution of ex-post operational costs and carbon footprint: proposed \ac{PPA} vs. \ac{ToU}.}
    \label{fig:study-case1-operational costs}
\end{figure}
In \cref{fig:study-case1-operational costs}, we show the empirical \acp{PDF} of the operational costs and carbon footprint across days (estimated with \ac{KDE} on the observed daily metrics). One can observe that using our method within the proposed \ac{PPA} shifts the costs \ac{PDF} to the left, suggesting a cost reduction in most cases. However, one can also observe that the right-side tail of the \ac{PDF} is extended, suggesting rare occasions of large energy supply costs (likely caused by extreme imbalance costs events). Interestingly, the footprint \ac{PDF} is shifted to the right, suggesting an increase in the footprint of the system. Intuitively, this can be attributed to less cost reduction opportunities in the baseline (since the tariffs are less dynamic), which lead to more opportunities of reducing carbon emissions without negatively impacting the costs. It is important to remember that the cost of carbon emissions is a parameter of the day-ahead bidding strategy (see \cref{sec:carbon-param}). 

To quantify the differences between the two supply agreements, we describe the \acp{PDF} in \cref{tab:tou-v-baseline-v-no-flex}. From ex-post results, we expect the custom supply agreement to lead to a reduction in expected operational costs of approximately 22\%, and a footprint increase of approximately 6\%. In the custom \ac{PPA} case, we observe an underestimation of the two metrics in their ex-ante forecasts, likely due to an underestimation of imbalance prices.

\begin{table*}[ht]
\caption{Quantitative Description of Operational Footprint and Cost}
\centering
\small
\setlength{\tabcolsep}{4pt}
\resizebox{\textwidth}{!}{%
\begin{tabular}{|l|*{16}{c|}}
\hline

Supply scheme & \multicolumn{4}{c|}{Ex-post costs (EUR)} & \multicolumn{4}{c|}{Ex-ante costs (EUR)}
& \multicolumn{4}{c|}{Ex-post emissions (kgCO$_{2}$eq)} & \multicolumn{4}{c|}{Ex-ante emissions (kgCO$_{2}$eq)}\\
\hline
- & Q25\% & Mean & Q75\% & $\sigma$ & Q25\% & Mean & Q75\% & $\sigma$ & Q25\% & Mean & Q75\% & $\sigma$ & Q25\% & Mean & Q75\% & $\sigma$ \\
\hline
ToU                & 171 & 197 & 216 & 31 & 169 & 195 & 219 & 32 & 79 & 98 & 105 & 28 & 73 & 91 & 105 & 24 \\
Custom             & 128 & 153 & 165 & 38 & 126 & 144 & 154 & 26 & 86 & 104 & 109 & 33 & 77 & 96 & 109 & 27 \\
Custom, no WL flex & 150 & 168 & 186 & 29 & 144 & 161 & 172 & 24 & 83 & 104 & 110 & 31 & 80 & 97 & 109 & 26 \\
\hline
\end{tabular}%
}
\label{tab:tou-v-baseline-v-no-flex}
\end{table*}
\subsubsection{Study Case II --- on the Value of Flexible \ac{WL} in the Spot Market}
\label{study-case:flex-value}
In this section, we quantify the economic value of flexible \ac{WL} in the day-ahead planning phase. We use the flexible demand scenarios for computational resources (i.e., \acp{CPU} and \acp{GPU}) generated in study case \ref{study-case:market-v-tou}. The daily flexible demand is uniformly distributed over the corresponding inelastic demand scenario, ensuring the total daily demand for computational resources remains unchanged. This allows us to isolate the impact of WL flexibility.

\begin{figure}[]
    \centering
    \includegraphics[width=.8\linewidth]{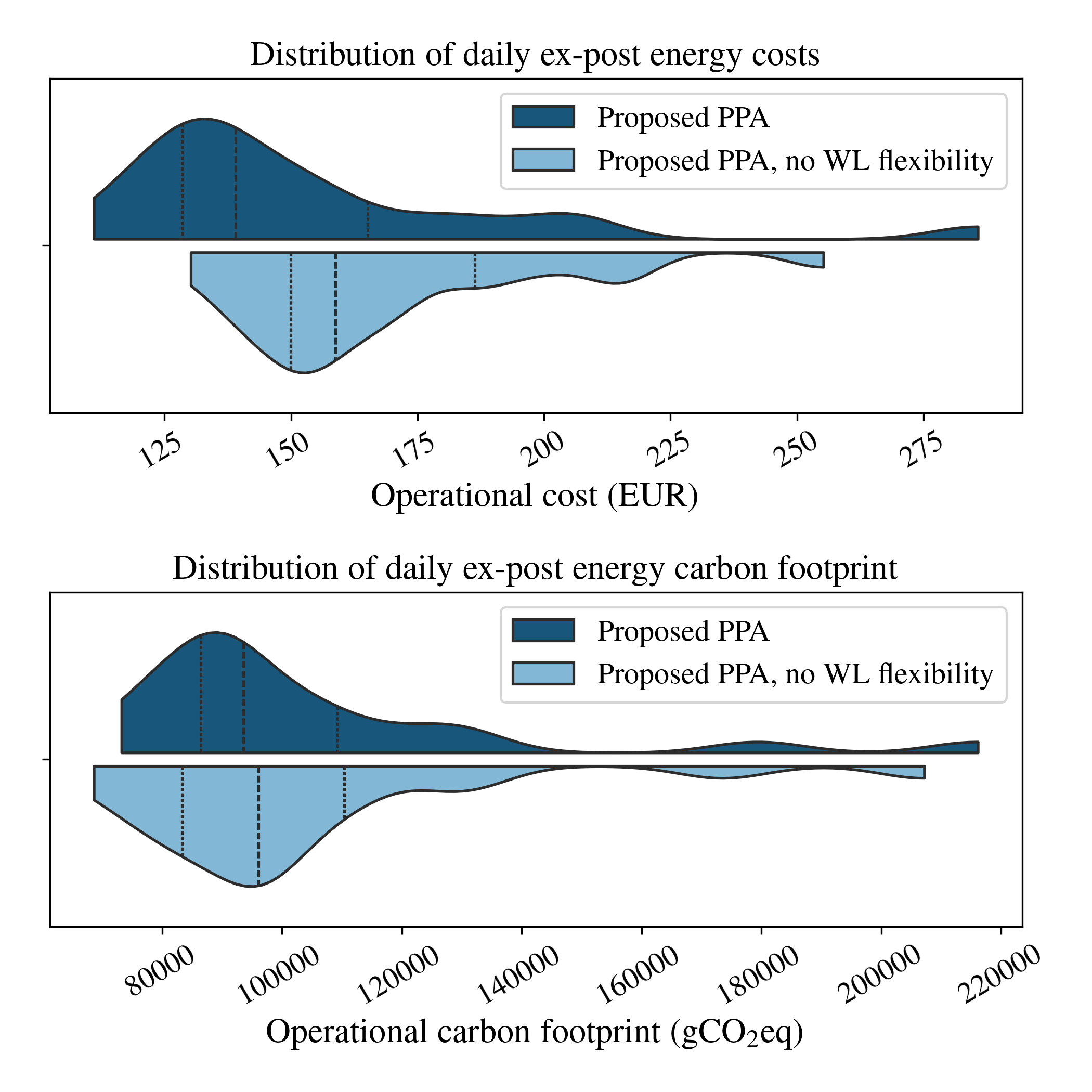}
    \caption{Ex-post energy supply operational costs and footprint: no \ac{WL} elasticity vs. 50\% of flexible \ac{WL}.}
    \label{fig:study-case2-operational costs}
\end{figure}

From \cref{fig:study-case2-operational costs} and \cref{tab:tou-v-baseline-v-no-flex}, we conclude that the expected cost reduction from flexible \ac{WL} is approximately \SI{15}{EUR} per day. Based on the scenarios analyzed, the average flexible demand amounts to about 6000 CPUh and 1500 GPUh per day. Using the pricing reported in \cite{epfl_operational_2024}, their corresponding value is about \SI{660}{EUR}. Thus, the observed savings represent less than 2.5\% of the total value of the \ac{WL}, assuming it were inelastic. This marginal reduction offers little economic incentive to users, making it unlikely that such discounts would encourage flexible resource bookings. 

These results suggest that, under current pricing schemes, \ac{WL} flexibility does not provide enough value in the spot market alone to motivate data center operators to propose flexible rates. However, other factors (such as system reliability, local policies, participation to other electricity markets, or environmental goals) could justify flexible billing schemes.
\subsubsection{Study Case III --- on the Impact of Virtual De-Rating}
\begin{figure}[ht]
    \centering
    \includegraphics[width=1\linewidth]{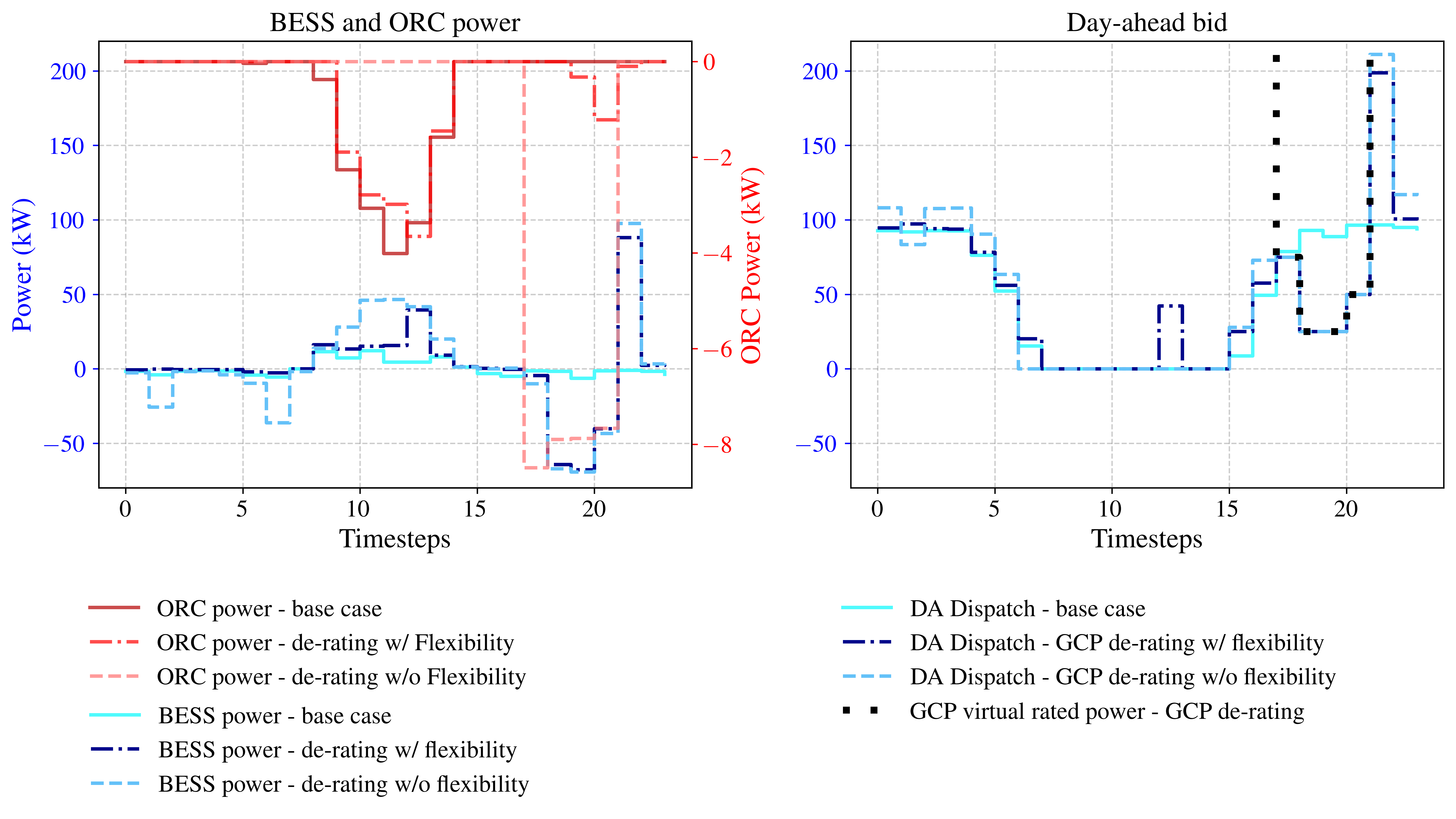}
    \caption{Dispatch for 18.07.2025, no de-rating vs. de-rating with and without \ac{WL} flexibility.}
    \label{fig:study-case3-dispatch-de-rating}
\end{figure}

We assess the impact of virtual power de-rating of the \ac{GCP} on day-ahead planning. The \ac{DSO} anticipates a period of high local demand between \SI{17}{\hour} and \SI{21}{\hour} and tries to alleviate system stress. In this regard, it issues a day-ahead request of virtual de-rating of the data center’s grid connection power. Specifically, we assume the nominal connection capacity of \SI{300}{\kilo\watt} to be temporarily reduced to:
\begin{itemize}
    \item \SI{75}{\kilo\watt} at 17:00,
    \item \SI{25}{\kilo\watt} at 18:00,
    \item \SI{25}{\kilo\watt} at 19:00,
    \item \SI{50}{\kilo\watt} at 20:00 (until 21:00).  
\end{itemize}
We evaluate the impact of de-rating on system planning in two scenarios: 50\% flexible \ac{WL} and no flexible \ac{WL}, and compare both to a baseline without virtual de-rating (with 50\%  flexible \ac{WL}). \Cref{fig:study-case3-dispatch-de-rating} shows the effect of the virtual de-rating on day-ahead power profiles. On the right-side of the figure, we observe that the de-rating constraint is satisfied in the day-ahead bid. In the left-side of the figure, we see that the \ac{BESS} and \ac{ORC} compensate most of the shortfall caused by the virtual power de-rating. Interestingly, the \ac{ORC} has a crucial role in the case without \ac{WL} flexibility, highlighting its contribution in highly constrained operating conditions (e.g., de-rating and no \ac{WL} flexibility)\footnote{The heat price selected in \cref{sec:heat-price} makes it generally more advantageous to deliver recovered heat to the district heating network rather than to the \ac{ORC} system, if the district heating network needs that heat.}. 
In the baseline, the expected energy supply cost is \SI{101}{EUR}. With virtual de-rating it increases by 10\% (i.e., +\SI{11}{EUR}) with workload flexibility and 33\% (i.e. +\SI{34}{EUR}) without. This increase is mainly caused by the additional aging induced on the \ac{BESS} and the reduced exports of heat. Given that this increase is similar to the daily savings enabled by the custom contract (as shown in \cref{tab:tou-v-baseline-v-no-flex}, larger than \SI{30}{EUR} in expectation), we conclude that the proposed contract is economically attractive for the data center, as long as the triplet $(\cs{P}{gcp,cap}{$t$},~\cs{t}{daily,lim}{},~\cs{t}{weekly,lim}{})$ of the \ac{PPA} are carefully selected.

\begin{figure}[ht]
    \centering
    \includegraphics[width=1\linewidth]{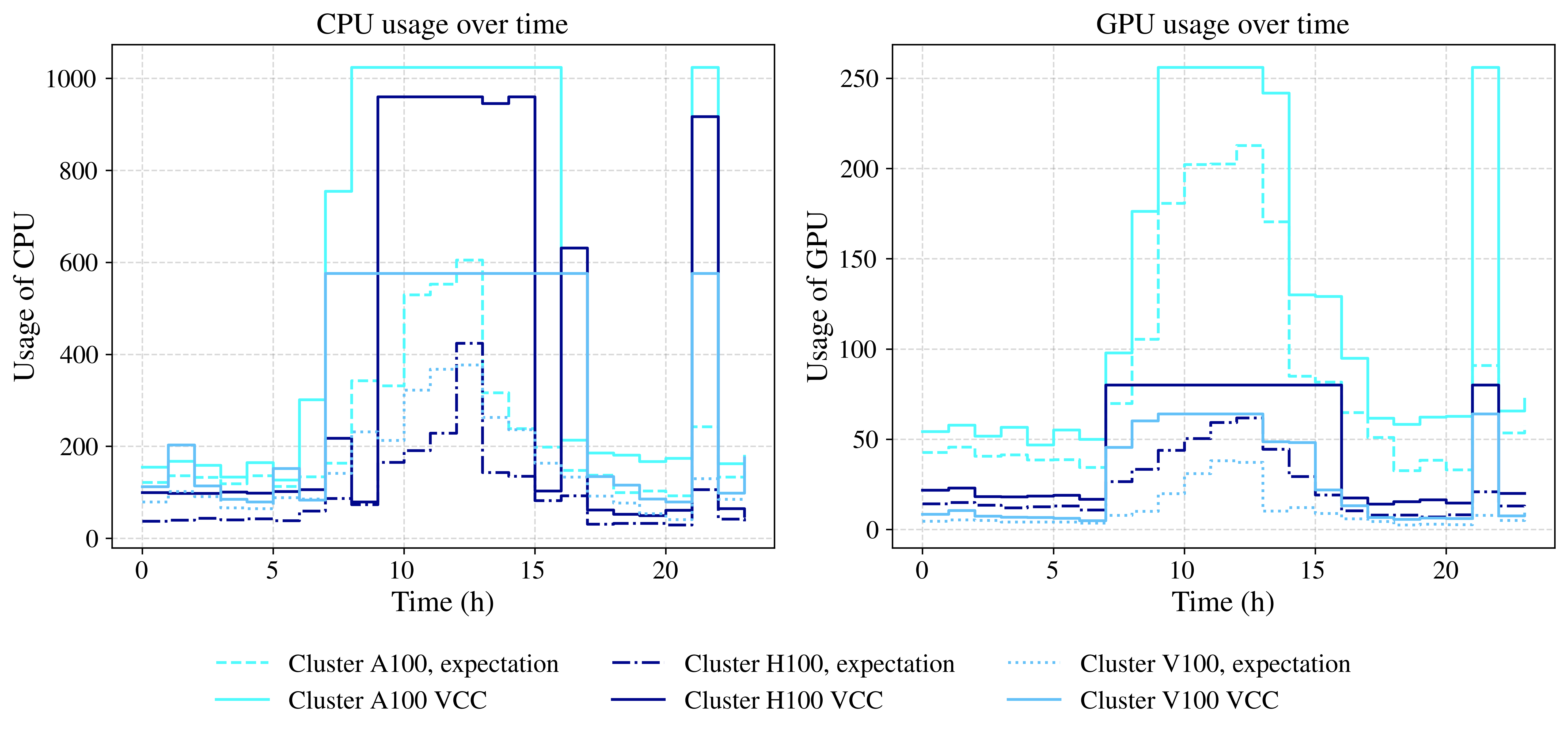}
    \caption{Resource usage and \ac{VCC} for 18.07.2025, de-rating with \ac{WL} flexibility.}
    \label{fig:study-case3-vcc-de-rating}
\end{figure}

If flexible workload is available, the bidding strategy leverages \acp{VCC}, as shown in \cref{fig:study-case3-vcc-de-rating}. Most \ac{WL} is scheduled during daytime hours, driven by high solar generation and low spot prices. In contrast, costly morning hours and periods of reduced grid capacity exhibit low virtual capacity, reflecting a low inclination to schedule \ac{WL} during these periods. During real-time operation, the workload scheduler should use the \acp{VCC} to avoid overscheduling \ac{WL} during constrained hours. 

\section{Conclusion}
In this work, we presented a practical risk-averse day-ahead bidding strategy for small-scale data centers gaining access to the spot electricity market through a custom \ac{PPA} with the local \ac{DSO}. The strategy accounts for imbalance costs and carbon emissions. It leverages flexible \ac{WL}, \ac{BESS}, local \ac{PV} generation, and waste heat recovery to minimize data center operational costs.

Through three study cases based on an academic datacenter in Lausanne, Switzerland, we analyzed the proposed \ac{PPA} combined with our bidding strategy. In the first study case, we compared the proposed market scheme to a classic \ac{ToU} billing scheme and showed that it is expected to reduce energy supply costs by 22\%, thus showing that there are opportunities for small-scale data centers in the day-ahead electricity market. The second study case assessed the value of \ac{WL} flexibility and showed that the marginal gains it enables are unlikely to motivate data centers to propose custom billing schemes for flexible \ac{WL} (from the energy supply perspective). Finally, the third study case showed that virtual \ac{GCP} transfer capacity de-rating requests by the \ac{DSO} should not significantly impact operational costs if the supply agreement is well-designed.

We believe this work can benefit both \acp{DSO} and \acp{DCO}, as the proposed bilateral billing scheme provides value to both parties. The main limitation of the study is its focus on the day-ahead scheduling phase, which requires simplifying assumptions, particularly in \ac{WL} modeling. Future work will extend this approach toward real-time control strategies for data centers, building on the bidding framework proposed here.

\section*{Acknowledgments}
This work is supported by the Heating Bits EPFL project financed by the Solutions4Sustainability initiative. Data on data center usage and power consumption was provided by the Research Computing Platform of EPFL.
\bibliographystyle{IEEEtran}
\bibliography{references}

%



\end{document}